\documentclass[prb,twocolumn,preprintnumbers,showpacs]{revtex4}
\usepackage{amssymb, amsmath, bbm}
\usepackage{graphicx}
\renewcommand{\narrowtext}{\begin{multicols}{2} \global\columnwidth20.5pc}

\newcommand{\s}{{\sigma}}

\def\be{\begin{eqnarray}}
\def\ee{\end{eqnarray}}
\newcommand{\nn}{\nonumber\\}
\newcommand{\Eq}[1]{Eq.~(\ref{#1})}

\newcommand{\cE}{ {\cal E} }

\newcommand{\cH}{ {\cal H} }

\newcommand{\bnormal}[1]{\colon\!{#1}\colon}

\newcommand{\sgn}{\text{sgn}}
\newcommand{\ket}[1]{\left|\,{#1}\right>}

\newcommand{\braket}[2]{\left<{#1}\,|\,{#2}\right>}
\newcommand{\ketphy}[1]{\left|\,{#1}\right>{\substack{\\\\\hspace{-.07cm}l}}}

\newcommand{\braketphy}[2]{{\substack{\\\\l\hspace{-.07cm}}}\left<{#1}\,|\,{#2}\right>{\substack{\\\\\hspace{-.07cm}l}}}

\begin{document}
\title{The Luther-Emery liquid: Spin gap and anomalous flux period}
\author{Alexander Seidel}
\affiliation{Department of Physics, University of California at Berkeley, Berkeley, CA 94720, USA}
\affiliation{Materials Sciences Division, Lawrence Berkeley National Laboratory, Berkeley, CA 94720, USA }
\author{Dung-Hai Lee}
\affiliation{Department of Physics, University of California at Berkeley, Berkeley, CA 94720, USA}
\affiliation{Materials Sciences Division, Lawrence Berkeley National Laboratory, Berkeley, CA 94720, USA }
\date{\today}
\begin{abstract}
We study the dependence of the ground state energy on an applied
Aharonov-Bohm flux $\Phi$ for the Luttinger model with large momentum
scattering.
Employing
the method of finite size bosonization, we show that for systems
with a spin gap but with gapless charge degrees of freedom, the
ground state energy has an exact period of $hc/2e$, i. e. {\em
half} a flux quantum, in the limit of large system size $L$. Finite
size corrections are found to vanish exponentially in $L$. This
behavior is contrasted to that of the spin gapless case, for both
even and odd particle number. Generalizations to finite
temperature are also discussed.
\end{abstract}
\maketitle

\section{Introduction}

Models of interacting electrons in one spatial dimension are very
valuable for the understanding of strongly correlated systems. This
is because there exist theoretical methods enabling us to
determine their physical properties reliably. Indeed, by combining
perturbative renormalization group\cite{SOLYOM}, bosonization, and
Bethe ansatz techniques, a wealth of interesting phases in one
dimension has been discovered.

While some properties of these phases are unique in one dimension,
others have their higher dimensional analogs. For example the
independent gapless spin- and charge- excitations and the
vanishing quasiparticle weight of the {\em Luttinger liquid}
\cite{HAL1} are unique in 1D. However, the fact that it has a
finite charge compressibility and Drude weight is analogous to a
normal metal in higher dimensions. As another example, like
systems in higher dimensions, a Mott insulating state is realized at
half filling for repulsive interactions. However,
the fact that antiferromagnetic long range order is absent and
that spin 1/2 excitation exists in the half-filled Mott state are
special features of 1D.

Furthermore, in one dimension there exists a phase, the {\em Luther-Emery
liquid} \cite{LE}, which exhibits a spin gap and no charge gap. In
addition, as in the {\em Luttinger liquid}, the DC electric
conductivity is infinite. The above characteristics suggest that
the {\em Luther-Emery liquid} is a 1D analog of a superconductor.
However, up until very recently an important question remained
unanswered: ``Do electrons pair in the {\em Luther-Emery liquid}
?'' The best way to answer that question is to determine whether
the magnetic flux period is $hc/e$ or $hc/2e$.\cite{BYERS, KOHN}
However, since the spin and charge degrees of freedom are
manifestly separate in the effective theory describing the {\em
Luther-Emery liquid}, and the vector potential enters only in the
charge action, it is difficult to see why the flux period
for a {\em Luttinger liquid} and a {\em Luther-Emery liquid} should be
different.
\\

In a recent paper we addressed these issues in the one-dimensional
$t$-$J$-$J'$-model in the limit of vanishing exchange couplings.
\cite{tJJflux} Fortunately, both a spin gapless phase as well as a
spin gapful phase appear in this limit.\cite{OGLURI} In Ref. 
\onlinecite{tJJflux}
we have demonstrated that while the flux period is $\Phi_0\equiv hc/e$ in the
former\cite{BUETTIKER}, it indeed becomes $hc/2e$ in the latter.
In particular we have shown that as a function of the
Aharonov-Bohm flux, the ground state energy of a spin gapped ring
is periodic with period $hc/2e$. Due to one-dimensionality the
energy barrier between adjacent minima is proportional to the
inverse circumference $L$ of the ring. For definiteness, we
therefore define the function \be\label{eq1}
  {\cal{E}}(\Phi)= \lim_{L\rightarrow \infty}\,L\,\bigl( E_0(\Phi)-E_0(0)\bigr)
\ee where $E_0(\Phi)$ is the ground state energy of the system as
function of flux.

Despite the above progress, the question ``do all {\em Luther-Emery
liquids} exhibit an $hc/2e$ flux period, and hence electron pairing
?'' remains to be answered. In this paper, we show that the answer
to the above question is indeed affirmative. Technically we start
from the Luttinger Hamiltonian with the $g_1$ channel
scattering.\cite{SOLYOM} We bosonize this model using the
constructive formalism \cite{HEIDENREICH, HAL6} which provides
rigorous operator identities on the Hilbert space of the finite
size system. We show that due to a set of constraints on the total
charge/spin number/current operators\cite{HAL1,HAL7}, the state of
the spin sector impacts the charge sector through a twisting of
the boundary condition. As a result, when the spin sector is
gapped by the large momentum transfer two body scattering, the
charge channel flux period becomes $hc/2e$.
\\

In the literature, the fact that there exist constraints on the
total charge/current operators in bosonization has been employed 
by Loss\cite{LOSS} for spinless fermion systems to study particle
number parity effects. Regarding spinful
fermions, Ref. \onlinecite{KUS} used a method similar to ours to
determine the flux period for the Hubbard model. However, the
author concluded that the flux period is always
$hc/2e$ regardless of whether a spin gap exists, which we believe to
be in error. 
Furthermore, a common reasoning encountered in the literature is to attribute
the $hc/2e$ flux period to the dominance of singlet superconducting
(SS) correlations at long distance and low energy, rather than to
the appearance of a spin gap. { It has, however, been noted that states with 
dominant charge density wave (CDW) correlations may also feature this
anomalous flux period (see, e. g., Refs. \onlinecite{tJJflux,SUDBO,STECHEL}). 
Here we argue that this is just the case when there
is a spin gap. In this case, it is natural to interpret the state as
being formed by Cooper pairs. The degree of coherence of
these pairs will determine if the state is more appropriately
thought of as CDW-like or SS-like on not too large length scales.
In this picture, one naturally expects the flux period to be 
one half of a flux quantum. In the following,
we will show that regardless of the correlation 
functions in the charge sector, the existence of a spin gap {\em alone}  
indeed causes the $hc/2e$ flux period in systems with even particle number.
}
\\

The structure of this paper is as follows: In section \ref{geology} 
we present the Luttinger model with large momentum scattering 
and state the
selection rules between charge and current quantum numbers that
characterize its Hilbert space. In section \ref{bosonize}
we briefly review the formalism of constructive bosonization
and introduce some notation.  In section \ref{proof} we complete
the proof that the flux period will be $hc/2e$ in the presence 
of a spin gap, and contrast this behavior with that expected
in the spin gapless case for even and odd particle number. We will 
also comment on finite temperature effects here. Our conclusions 
are summarized in section \ref{conclusion}. Appendix \ref{appA}
discusses the finite size refermionization of the spin
part of the Hamiltonian, supplementing our line of arguments
given in the bulk of this paper. Appendix \ref{appB} is devoted
to the use of conjugate phase variables in the construction 
of Klein factors.

\section{The model and the selection rules\label{geology}}

\begin{figure}[t]
\includegraphics[width=5cm,height=5cm]{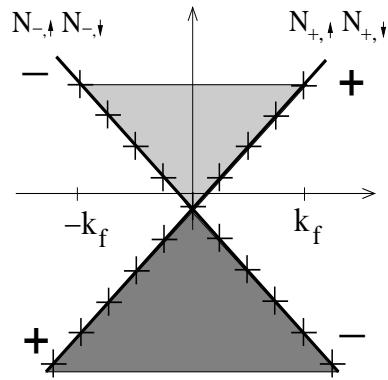}
\caption{\label{TL} Right- and left-moving branches of the
Luttinger model. The crosses denote the allowed momenta $k=2\pi n/L$
for periodic boundary conditions. The dark shaded region
represents the occupied momentum states in the ``vacuum''
$N_{r,s}=0$. The vacuum chemical potential lies between the last
occupied and first unoccupied states as indicated by the
horizontal line. The light shaded region corresponds to a
different filling. The Fermi momentum corresponding to the 
latter is given by $k_f=2\pi N_{r,s}/L$.}
\end{figure}

The Tomonaga-Luttinger Hamiltonian describes a gas consisting of
right and left moving chiral fermions, each suffering small-momentum
transfer scattering in a one-dimensional system of size $L$: \be
 &&H_{TL}=H_0+H_2+H_4 \label{HTL}\\
 &&H_0=\sum_{r,k,s}(r v_Fk-\mu):c^\dagger_{rks}c_{rks}:\label{H0}\\
 &&H_2=\;\frac{1}{ L}\sum_{q,s,s'}\left(g_{2||}\delta_{s,s'}+g_{2\perp}\delta_{s, -s'}\right)\rho_{+,s}(q)\rho_{-,s'}(-q)\label{H2}\\
  &&H_4=\frac{1}{2L}\sum_{r,q,s,s'}\left(g_{4||}\delta_{s,s'}+g_{4\perp}\delta_{s, -s'}\right):\rho_{r,s}(q)\rho_{r,s'}(-q):\nn
&&\label{H4} \ee Here $k=2\pi n/L$ denotes the allowed momenta
under periodic boundary condition, the fermion 
operator $c_{rks}$ annihilates a right ($r=+$) or left ($r=-$)
moving fermion with momentum $k$ and spin $s$ (see Fig. \ref{TL}),
$\mu=v_F\pi/L$, and \be :{\cal O}:\,\equiv {\cal O} - \left< {\cal
O}\right>_0.\label{fnormal}\ee
In the above, $<...>_0$ denotes the expectation value taken in the
vacuum state defined as the ground state of \Eq{H0}. The density
operators appearing in Eqs. (\ref{HTL})-(\ref{H4}) are defined as
\be\label{rhors}
 \rho_{r,s}(q)\equiv\sum_k :c^\dagger_{r,k+q,s}c_{r,k,s}:
\ee The $q\!=\!0\,$-component of these operators, \be
N_{r,s}\equiv \rho_{r,s}(0) \ee measures the extra number of
$(r,s)$ type fermions added on top of the vacuum. All four integers
$N_{r,s}$ are conserved by $H_{TL}$. These quantum numbers play an
important role in the rest of the paper. Their importance in the bosonization procedure 
has been stressed by Heidenreich et al.\cite{HEIDENREICH} and Haldane\cite{HAL6}.

Out of the four operators $N_{r,s}$ we can form the following
linearly independent number and current operators:
\be\label{qnumbers} &N_{\rho}=\sum_{r,s}\,N_{r,s}\quad,\quad
J_{\rho}=\sum_{r,s}\,r\,N_{r,s}\nn
&N_{\sigma}=\sum_{r,s}\,s\,N_{r,s}\quad,\quad
J_{\sigma}=\sum_{r,s}\,rs\,N_{r,s}, \ee
 where the indices $\rho$
and $\sigma$ stand for charge and spin respectively. 
It will be important in the following to note that 
in any one-band model with single particle states 
symmetrically occupied between $k_f=2\pi N_{r,s}/L$ and $-k_f$,
the total particle number is actually given by 
\begin{equation}
  N=2+\sum_{r,s}\,N_{r,s}=2+N_\rho
\end{equation}
The reason for this is that the states at $k=0$,
which consist of 4 degenerate states in the Luttinger
model rather than 2, have not been included in the
definition of the $N_{r,s}$ (see Fig. \ref{TL}).

There are important relations between the integer quantum numbers
defined in \Eq{qnumbers}. For example $N, N_{\rho,\s},
J_{\rho,\s}$ are either all odd or all even. In addition, the
average of $N_{\rho}$ and $J_\rho$ has the same even-odd parity as
the average of $N_\s$ and $J_\s$, while they both have opposite
even-odd parity as the average of $N$ and $J_\rho$. These
constraints are summarized by the following ``selection
rules''\cite{HAL7}: 
\begin{subequations}\label{sel}
\begin{align}
&(-1)^N=(-1)^{N_\rho}=(-1)^{J_\rho}=(-1)^{N_\sigma}=(-1)^{J_\sigma}\label{sel1}\\
&-(-1)^{(N+J_\rho)/2}=(-1)^{(N_\rho+J_\rho)/2}=(-1)^{(N_\sigma+J_\sigma)/2}\label{sel2}
\end{align}
\end{subequations}
which follow from the definitions \Eq{qnumbers} and the fact that 
the $N_{r,s}$ are integer.
For most of the paper, we shall primarily concentrate on the
case where $N$ is even. While selection rule \Eq{sel1} then
requires the same of all the other quantum numbers, 
it is the selection rule \Eq{sel2}
that imposes a coupling between the spin and charge quantum numbers
which ultimately determines the value of the flux period.
\\

The Tomonaga-Luttinger Hamiltonian \Eq{HTL} is exactly solvable.
\cite{LIEBMATTIS, HEIDENREICH, HAL6} The solution describes a
system with gapless spin and charge excitations. A spin gap may be
opened by the addition of the following large-momentum transfer
scattering term\cite{LE}: 
\be
H&=&H_{TL}+H_1\label{g-ol}\\
H_1&=&H_{1,||}+H_{1,\perp}\nn
&=&-\frac{1}{L}\sum_{k,k',q,s,s'}(g_{1||}\delta_{ss'}+g_{1\perp}\delta_{-s
s'})\nn&\times&
:c^\dagger_{+,k'+q,s'}c_{+,k,s}::c^\dagger_{-,k-q,s}c_{-,k',s'}:\label{H1}
\ee 
When the number of particles is incommensurate with the number
of lattice sites, \Eq{g-ol} is the generic Hamiltonian including
the most relevant two-body scattering terms. The inclusion of
$H_1$ destroys the exact solubility of the model, and
at the same time it destroys the conservation of $J_\sigma$.
However since $H_1$ changes $J_\sigma$ in multiples of $4$, the
parity $(-1)^{J_\sigma/2}$ remains conserved. As a result the
selection rules \Eq{sel1}, \Eq{sel2} remain valid even in
the presence of $H_1$.

\section{\label{bosonize}Bosonization}

Under suitable choices of parameters, \Eq{g-ol} can describe a
translationally invariant system of spin-$1/2$ fermions with a
spin gap but no charge gap, i.e., a {\em Luther-Emery liquid}.
In the rest of the paper we study the dependence of the ground
state energy of such a model as a function of an applied
Aharonov-Bohm flux. Technically we employ the constructive
bosonization method\cite{HEIDENREICH, HAL6} extensively reviewed
in Refs. \onlinecite{DELFT} and \onlinecite{VOIT}. In the
following we shall just summarize the main bosonization rules.

Due to the following commutation relation between the density
operators \be [\rho_{r,s}(-rk),\rho_{r',s'}(r'k')]={kL\over
2\pi}\delta_{rr'}\delta_{ss'}\delta_{k,k'}.\ee we define boson
creation operators for each momentum $q\neq 0$ 
and each spin $s$: \be
&&b^\dagger_s(q)=\sqrt{\frac{2\pi}{|q|L}}\,\sum_r\,\Theta(rq)\,\rho_{r,s}(q)\nn
&&[b_s(q),b^\dagger_{s'}(q')]=\delta_{s,s'}\delta_{q,q'},~~[b_s(q),b_{s'}(q')]=0,\ee
where $\Theta(x)$ is the Heaviside step function. The bosonization
of the local fermion operators \be
\psi_{r,s}(x)=\frac{1}{\sqrt{L}}\sum_k e^{ikx}c_{r,s}(k) \ee then
proceeds by means of the introduction of a non-Hermitian
bosonic field \be\label{varphi}\varphi_{r,s}(x)=-\frac{\pi
rx}{L}N_{r,s}+i\sum_{q\neq
0}\sqrt{\frac{2\pi}{L|q|}}\Theta(rq)e^{iqx-rq\alpha/2}b_s(q),\nn
\ee in terms of which the fermion creation operators can be
written\cite{HEIDENREICH, HAL6, DELFT} as:\be
\psi^\dagger_{r,s}(x)=\frac{1}{\sqrt{L}}\,A_{r,s}\,e^{\,i\varphi^\dagger_{r,s}(x)}\,e^{\,i\varphi_{r,s}(x)}\,
e^{\,i\overline\varphi_{r,s}},\label{bosid}\ee where the 
factor \be
A_{r,s}=e^{i\frac{\pi}{2}\left(r\sum_{s'}N_{-r,s'}+s\sum_{s'}N_{r,s'}\right)},\label{Ars}\ee
is introduced to ensure the proper anticommutation relations
between the fermion operators (\ref{bosid}) carrying different $r$
and $s$. It commutes with all the spatially dependent fields in \Eq{bosid}.
A positive infinitesimal $\alpha$ was introduced in
\Eq{varphi} to ensure the convergence of commutators between
operators. The operator $\overline\varphi_{r,s}$ is conjugate to
$N_{r,s}$,
\be\left[\overline\varphi_{r,s},N_{r,s}\right]=i\label{comm0}.\ee
{
Note that the validity of
\Eq{comm0} formally requires $N_{r,s}$ to have a continuous spectrum (see Appendix \ref{appB}). 
This is clearly not the case in the physical Hilbert space ${\cal H}_{phys}$ we have been working in so far. We find it convenient, however, to introduce a larger Hilbert space $\cal H$, where the $N_{r,s}$ operators have a continuous 
spectrum. This construction is analogous to the embedding of a discrete lattice into a continuous space, and is
reviewed in Appendix \ref{appB}. 
 To ensure that the Hamiltonian, as well as physical observables, do not lead out of ${\cal H}_{phys}$, the operators $\overline\varphi_{r,s}$ may only enter through integer powers of the unitary operators $\exp(i\overline\varphi_{r,s})$, which raise the $N_{r,s}$ by 1. We shall have occasion though, e. g. in Appendix \ref{appA}, to work in a larger subspace of $\cal H$ defined below. 
Formally, it is most convenient to define operators that are valid everywhere in $\cal H$. 
The anticommuting operators $A_{r,s}\exp(i\overline\varphi_{r,s})$ are also 
known as {\em Klein factors} in the literature. 
}

It is customary to further define local
Hermitian fields each associated with the spin ($\s$) or charge
($\rho$) degrees of freedom,  \be
\label{phi}&&\phi_{\rho,\sigma}(x)
=\frac{1}{4}\,\sum_r\,r\left(\varphi_{r,\uparrow}(x)\pm\varphi_{r,\downarrow}(x)+\text{h.c.}
\right) + \overline\phi_{\rho,\sigma}\nn &&\overline
\phi_{\rho,\sigma}
=\frac{1}{4}\,\sum_r\,r\left(\overline\varphi_{r,\uparrow}\pm\overline\varphi_{r,\downarrow}
\right), \ee as well as their ``dual'' fields, \be\label{theta}
&&\theta_{\rho,\sigma}(x)=\frac{1}{4}\,\sum_r\,
\left(\varphi_{r,\uparrow}(x)\pm\varphi_{r,\downarrow}(x)+\text{h.c.}
\right)+ \overline\theta_{\rho,\sigma}\nn
&&\overline\theta_{\rho,\sigma}=\frac{1}{4}\,\sum_r\,\left(\overline\varphi_{r,\uparrow}\pm\overline\varphi_{r,\downarrow}
\right) \ee Written in terms of the spin and charge boson
operators \be\label{mode} b_{\rho,\sigma}(q)=\frac{1}{\sqrt
2}\left(b_{\uparrow}(q)\pm b_{\downarrow}(q)\right)\text{,} \ee
the above local fields read ($\nu=\rho,\sigma$) \be\label{modeex}
\phi_\nu(x)&=&\overline\phi_\nu-\frac{\pi}{2}\frac{N_\nu x}{L}-\frac{i}{2}
\sum_{q\neq 0}\sgn(q)\sqrt{\frac{\pi}{L|q|}}\nn&\times&e^{-iqx-|q|\alpha/2}\left(b^\dagger_\nu(q)+b_\nu(-q)\right)\label{phiexp}\\
\theta_\nu(x)&=&\overline\theta_\nu-\frac{\pi}{2}\frac{J_\nu
x}{L}-\frac{i}{2}\sum_{q\neq
0}\sqrt{\frac{\pi}{L|q|}}\nn&\times&e^{-iqx-|q|\alpha/2}\left(b^\dagger_\nu(q)-b_\nu(-q)\right).\label{thetaexp}
\ee From Eqs. \eqref{phiexp} and \eqref{thetaexp} it is evident that
$N_\nu$ and $J_\nu$ are the winding numbers of $\phi_\nu$ and
$\theta_\nu$ respectively, and $\overline\phi_\nu$ and
$\overline\theta_\nu$ are the spatial averages of $\phi_\nu$ and
$\theta_\nu$. It is simple to check that $\overline\phi_\nu$ and
$\overline\theta_\nu$ are the conjugate operators of $J_\nu$ and
$N_\nu$ respectively, i.e., 
\be\label{comm1}
 \left[\overline\phi_{\nu},J_\nu\right]=i\;,\;\left[\overline\theta_{\nu},N_\nu\right]=i.
\ee
{Note that although the commutation relations \eqref{comm1} are analogous to
\Eq{comm0}, the operators  $e^{i\overline\phi_\nu}$ and
$e^{i\overline\theta_\nu}$ lead out of the physical subspace.
This is so since within ${\cal H}_{phys}$, the quantum numbers 
$N_\nu$, $J_\nu$ cannot be raised or lowered by 1 independently, but
are subjected to the selection rules \Eq{sel}.  Within this space,
only powers of $e^{4i\overline\phi_\nu}$
and $e^{4i\overline\theta_\nu}$ are allowed.\cite{NOTE1}
However, within the larger space $\cH$ introduced above, the operators 
 $e^{i\overline\phi_\nu}$ and $e^{i\overline\theta_\nu}$ are
nonetheless well defined objects. It is convenient to introduce
a space of ``fractional'' excitations, $\cH_{frac}$, generated
by acting on $\cH_{phys}$ with all possible combinations of
$e^{i\overline\phi_\nu}$, $e^{i\overline\theta_\nu}$. Within
$\cH_{frac}$, the quantum numbers $N_\nu$, $J_\nu$ are independent
integers. We must bear in mind, though, that all physically 
acceptable states live in $\cH_{phys}$.}

The inclusion of the zero modes $\bar{\phi}_\nu$
and $\bar{\theta}_\nu$ in Eqs. \eqref{phiexp} and \eqref{thetaexp}
ensures the proper commutation relations of these fields when the
system size $L$ is finite \be\label{comm2}
 \left[\phi_\nu(x),\theta_{\nu'}(x')\right]=i\frac{\pi}{4}\,\delta_{\nu,\nu'}\,\sgn(x-x').
\ee \Eq{comm2} suggests that the conjugate operator of
$\phi_\nu(x)$ is proportional to $\partial_x\theta_\nu(x)$, i.e,
\be\label{comm3}
&&\Pi_\nu(x)=-\frac{2}{\pi}\,\partial_x\,\theta_{\nu}(x)\nn
&&\left[\phi_\nu(x),\Pi_{\nu'}(x')\right]=i\,\delta_{\nu,\nu'}\,\delta(x-x').
\ee Similarly the conjugate operator of $\theta_\nu(x)$ is
proportional to $\partial_x\phi_\nu(x)$, i.e., \be\label{comm4}
&&\nu(x)=-\frac{2}{\pi}\,\partial_x\phi_{\nu}(x)\label{dens}\\
&&\left[\theta_\nu(x),\nu'(x')\right]=i\,\delta_{\nu,\nu'}\,\delta(x-x').
\ee
The physical spin or charge density is given by \Eq{dens}, and
in the absence of an applied flux, the physical (spin or
charge) current density is given by 
\be\label{current} j_\nu(x)=K_\nu v_\nu \Pi_\nu(x) =
-\frac{2}{\pi}K_\nu v_\nu\,\partial_x\,\theta_{\nu}(x), \ee 
which follows from the bosonized Hamiltonian given below. 
In the above expressions 
\be
&&v_\nu=\sqrt{\left(v_F+\frac{g_{4\nu}}{\pi}\right)^2-\left(\frac{g_{2\nu}}{\pi}\right)^2}\nn
&&K_\nu=\sqrt{\frac{\pi v_F -g_{2\nu}+g_{4\nu}}{\pi v_F
+g_{2\nu}+g_{4\nu}}}\nn &&g_{2\rho,\sigma}=\frac{g_{2||}\pm
g_{2\perp}-g_{1||}}{2}\nn &&g_{4\rho,\sigma}=\frac{g_{4||}\pm
g_{4\perp}}{2} \;.\ee
Since $\Pi_\rho$ is just the density of right moving fermions
minus the density of left moving fermions, it is appropriate
to interpret the coefficient $K_\rho v_\rho\equiv v_F^*$
in \Eq{current} as the renormalized Fermi velocity of the system.

In terms of $\phi_\nu(x)$ and $\theta_\nu(x)$ the  bosonization
identity \Eq{bosid} reads  \begin{equation}
  \psi^\dagger_{r,s}(x)=\frac{1}{\sqrt{L}}\,A_{r,s}\,:e^{i\bigl(\theta_\rho(x)+r\phi_\rho(x)
  +s\bigl(\theta_\sigma(x)+r\phi_\sigma(x)\bigr)\bigr)}:\label{bosid2}
\end{equation} Here, $:():$ denotes boson normal ordering: all powers of
the fields $\varphi_{r,s}^\dagger$ are to be moved to the left of
powers of the fields $\varphi_{r,s}$, whereas positive powers of
the operator $\exp(i\overline\varphi_{r,s})$ are to appear on the
very right, and negative powers of the same operator are to appear
on the very left of the expression.

By means of \Eq{bosid2} the selection rules \Eq{sel} become
equivalent to the requirement
\be
  \psi_{r,s}(x)=\psi_{r,s}(x+L)\qquad \forall\, r,s.
\ee
This clearly illustrates the topological origin of these rules.
\\

We may now write the Hamiltonian \Eq{g-ol} entirely in terms of
the bosonic fields introduced above. The Tomonaga-Luttinger part
of the Hamiltonian, including the large momentum 
scattering term with parallel spin, 
takes the following quadratic form:
\begin{align}
  H_{TL}&+H_{1,||}\nn
      &=\sum_{\nu}\,v_\nu\left\{\sum_{q\neq 0}|q|\, \tilde b_\nu^\dagger(q)\tilde b_\nu(q)+\frac{\pi}{4L}\left(N_\nu^2/K_\nu+J_\nu^2\,K_\nu\right)\right\}\label{HTLdiag}\\
  &=\sum_\nu\,\frac{v_\nu}{\pi}\,\int dx\,\bnormal{\left\{K_\nu\bigl(\partial_x\theta_\nu(x)\bigr)^2+\frac{1}{K_\nu}\bigl(\partial_x\phi_\nu(x)\bigr)^2\right\}}\label{HTL2}
\end{align}
where the operators $\tilde b_\nu(q)$ are related to those in
\Eq{mode} by a Bogoliubov transformation. The large
momentum-transfer scattering with antiparallel spin term becomes \cite{LE}:
\begin{equation}\label{H12}
 H_{1,\perp}=-\frac{2g_{1\perp}}{L^2}\,\int dx\,\bnormal{\cos\bigl(4\phi_\sigma(x)\bigr)}
\end{equation}
Note that the coefficient of $4$ in the argument of the cosine
assures that the operator does not go out of the physical
subspace, as explained above.

The weak coupling renormalization group flow of the system
\Eq{g-ol} is well known\cite{CHUILEE,SOLYOM}: For $K_\sigma<1$,
the operator $H_{1,\perp}$ is relevant and a spin gap will be opened. This
is the case we will focus on in the following. For spin $SU(2)$
invariant systems  $g_{i||}=g_{i\perp}\equiv g_i$. In that case
$K_\sigma<1$ requires $g_{1}<0$, as discussed by Luther and
Emery.\cite{LE}

\section{The flux period\label{proof}}
\subsection{\label{gapped}The spin gapped case}

By virtue of Eqs. \eqref{HTL2} and \eqref{H12}, the model
\Eq{g-ol} takes the form \be\label{g-ol2}
   H=H_{TL}+H_1\equiv H_\rho+H_\sigma
\ee
where $H_\rho$ and $H_\sigma$ act exclusively on charge- and
spin- degrees of freedom respectively. The eigenstates are thus
direct products of charge states and spin states \be\label{factor}
   \left|c\right>\otimes\left|s\right>,
\ee and the ground state energy is the sum of spin and charge
energies \be\label{energy}
  E_0=E_0^c+E_0^s.
\ee
When $H_1$ causes a spin gap to open up, the spin sector of the
model \Eq{g-ol} is described by a gapped sine-Gordon field
theory.

In the following we shall focus on the $N_\sigma=0$ sector, which
is where the gapped spin ground state lies. In this sector
$H_{1,\perp}$ is relevant, and one may interpret the cosine term in
\Eq{H12} as a steep potential experienced by $\phi_\sigma(x)$. In
the limit of infinite system size where true symmetry breaking is
possible, one may think of $\phi_\sigma(x)$ as being locked to one
of the minima of the cosine potential. 
When this happens $\overline\phi_\sigma$, the spatial average of
$\phi_\nu(x)$, will 
take a c-number value equal to the respective minimum value of
$\phi_\sigma$. { At first, let us neglect the selection rules \Eq{sel}.
That is, we start by looking at the problem in the space $\cH_{frac}$ 
introduced in Section \ref{bosonize}, where in particular 
$J_{\sigma}$ is an independent integer valued
quantum number. Then we may regard the conjugate variable $\overline\phi_\sigma$ as an angular
variable with period $2\pi$. This notion becomes precise if we identify
 $\overline\phi_\sigma$ with its ``lattice version'' discussed in Appendix
\ref{appB}, which we shall do for the present purpose.\cite{NOTE9}} Within $[0,2\pi)$ there are 
four inequivalent minima of the cosine term in \Eq{H12}, and the
corresponding ground states in the spin sector can be labeled as
\be\label{philock}
\left|0\right>,~~\left|{\pi}/{2}\right>,~~\left|\pi\right>,~~\left|{3\pi}/{2}\right>,\label{fgd}\ee
where \be \overline\phi_\sigma \left|\phi\right> =
\phi\left|\phi\right>.\ee
As discussed earlier, the operator \be\hat\eta\equiv
(-1)^{J_\sigma/2}\equiv\exp\left(-i{\pi}J_\sigma/2\right)\ee commutes
with $H_\s$, hence its eigenvalues can be used to classify the spin
ground states. Unfortunately the states given in \Eq{fgd} are
not eigenstates of $\hat\eta$. Following Appendix \ref{appB},
it is easy to show that
\be \overline\phi_\sigma\,
\hat\eta\left|\phi\right>=\left(\phi+\frac{\pi}{2}\right)\hat\eta\left|\phi\right>,
\ee
where the eigenvalue on the right hand side is to be
understood modulo $2\pi$. 
We may hence choose the global phases in
\Eq{philock} such that \be
   \left|z\pi/2\right>=\hat\eta^z\left|0\right>.
\ee It is thus easy to form linear combinations \be
\left|\eta\right>=\sum_{z=0}^3 \eta^{-z}
\,\hat\eta^{z}\left|0\right> \ee such that \be
\hat\eta\left|\eta\right>=\eta\left|\eta\right>.\ee
We are now in a position to enforce the selection rules \eqref{sel}.
Given $N_\s=0$, the selection rule \Eq{sel1} requires  
$J_\sigma$ to be even. 
As a result only $\eta=\pm 1$ are allowed. 
We label these two states by
\be \label{pm}\left|+\right>,~~\left|-\right>.\ee Thus actually,
the ground state is only two-fold degenerate.
This degeneracy becomes further lifted in the case of a finite
system size $L$, to be discussed next.

For finite $L$, the notion that the field $\phi_\sigma$ is locked
to a classical value is no longer valid. In fact 
for finite $L$, even $\overline\phi_\sigma$ is subjected to quantum
fluctuations. This is explicit in \Eq{HTLdiag}, where the variable
conjugate to $\overline\phi_\sigma$, namely $J_\sigma$, enters the
Hamiltonian when $L$ is finite. Thus the spin ground state can
no longer be thought of as one of 
the ``locked'' spin states given by \Eq{philock}. 
 On the other hand, since $\eta$ remains a good quantum number, the
states in \Eq{pm} are still well defined as the respective ground
states in the $\eta=\pm$ sectors of the spin Hilbert space. We
note that the spin states in \Eq{pm} thus defined are
not strictly degenerate for finite $L$. It is important to
observe, however, that the difference in energy between these two
states vanishes exponentially in the system size $L$. One way to
see this is the well known fact that the gapped sine-Gordon field
theory is the low energy effective theory of a dimerized
spin-$1/2$ chain.\cite{HAL3} Here, the $\left|\pm\right>$ are
respectively the symmetric and antisymmetric combination of the
two dimer patterns. Since the two dimer patterns differ by a
macroscopic number of degrees of freedom, the tunnel splitting
between these two states should vanish exponentially 
with the system size. A slightly more direct way to see the above
is offered by the 
well known mapping between the gapped sine-Gordon theory and the
massive Thirring model.\cite{LE,COLEMAN} We will elaborate on
this point in Appendix \ref{appA}. The advantage of this method is that
at the special Luther-Emery point, it allows us to 
study the effect of a finite temperature.

For the purpose of this paper we may ignore the above
exponentially small energy difference between the states
\Eq{pm}. This is because such a tiny difference 
will drop out of in the limit taken in \Eq{eq1}. In this sense we
may still speak of a degeneracy in the spin sector of the model,
and regard the spin contribution $E_0^s$ in \Eq{energy} as
essentially independent of $\eta$ in the spin gapped case.

Naively the spin degeneracy discussed above seems to suggest that the
ground state of the full Hamiltonian \Eq{g-ol2} is degenerate.
However this is not so, and the reason for this is the selection rule \Eq{sel2}.
To demonstrate that let us assume the total particle number to be
$N=4m+2$, 
whereas $N_\sigma=0$.
According to the selection rule \Eq{sel2} the spin states
$\left|\pm\right>$ may not be combined with the same charge state.
The spin state $\left|-\right>$
may only be combined with a charge state whose current quantum
number $J_\rho$ is an odd multiple of $2$ and hence non-zero. The
presence of a non-zero current will cost an energy of order
$v_F^*/L$ as is evident from \Eq{HTLdiag}. The state
$\left|+\right>$, on the other hand, may be combined with a charge
state of zero current, which minimizes the charge energy. As a
result there is an energy splitting $\sim 1/L$ between the 
lowest energy state in the $\eta=+$ and $\eta=-$ sectors. 
We note that an analogous result was discussed by
Haldane\cite{HAL7} for the case of a vanishing spin gap and a
finite charge gap at commensurate band fillings. In contrast, here
we are interested in the effect of an applied Aharonov-Bohm (AB)
flux, which is of interest only when the charge sector is gapless.

{ The coupling to a vector potential $A(x)$ is determined by 
gauge invariance and can be worked out from the minimal
coupling requirement. We only consider the constant
vector potential $A(x)=\Phi/L$ corresponding to an AB flux.}
The correct coupling to $\Phi$ then follows from the formal replacement\cite{BYERS}
\be\label{ABflux}
 \psi_{r,s}^\dagger(x)\,\longrightarrow\, e^{\,-i\frac{2\pi}{L}\frac{\!\Phi}{\Phi_0}x}\,\psi_{r,s}^\dagger(x)
\ee
in the Hamiltonian, where a charge $-e$ is assumed. Here, the boundary conditions of the field $\psi_{r,s}^\dagger(x)$ remain the same, while the right hand side of \Eq{ABflux} will in general satisfy different boundary conditions. 
By \Eq{bosid2}, this is equivalent to the following replacement in
the Hamiltonian \Eq{HTL2} and the current \Eq{current}:
\be\label{mc} &&\theta_\rho(x)\,\longrightarrow\,
\theta_\rho(x)-\frac{2\pi}{L}\frac{\!\Phi}{\Phi_0}\,x \ee 
or, by \Eq{thetaexp}, simply 
\be\label{replace}
 J_\rho \longrightarrow\, J_\rho+4\,\frac{\!\Phi}{\Phi_0}\,.
\ee 
{
 Note that we did not attempt to introduce the gauge flux prior
to bosonization. This is due to the fact that the fermionic
field theory \eqref{HTL} suffers from the well known chiral anomaly.
The latter renders the global current of the model ambiguous in the presence
of a general AB flux, unless gauge invariance is manifestly
enforced. Through the ``Lenz rule'' $I=-c\,\partial E_0(\Phi)/\partial\Phi$ (Kohn, Ref. \onlinecite{KOHN}, see also \Eq{current2}), this ambiguity
also enters the ground state energy dependence on flux.
To deal with this problem will in any case require the use of
 gauge invariance, and this is most conveniently achieved in the 
final, bosonic 
language. It would be interesting to obtain Eqs. \eqref{mc}, \eqref{replace} 
via a more ``microscopic'' route, i. e. via bosonization of a microscopic
(lattice) Hamiltonian with flux; this is subject to current investigations.
We stress again, however, that \Eq{mc} is uniquely
determined by the minimal coupling principle.
}

\begin{figure*}[t]
\samepage
\begin{center}
\vspace{.0cm}
\includegraphics[width=5in]{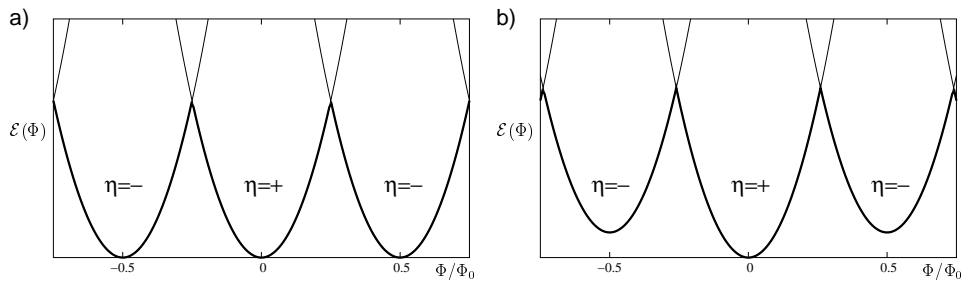}
\caption{\label{branches} Energy branches as function of flux for even total particle number
$N$, with and without spin gap. ${\cal E}(\Phi)$ is given by the lower envelope. The
alternating label $\eta=\pm$ describes the spin state corresponding to each branch, where
$N=4m+2$ is assumed. a) Spin gapped case. The flux period is $\Phi_0/2$. b) No spin gap. The
$\Phi_0/2$ flux periodicity is destroyed by a relative shift between the $\eta=+$ and
$\eta=-$ branches.}
\end{center}
\vspace{-.0cm}
\end{figure*}

From Eqs.  \eqref{HTLdiag}, \eqref{replace} the energy versus flux function ${\cal E}(\Phi)$ in  \Eq{eq1} is given
by 
\be\label{result}
 {\cal E}(\Phi)= \min_{J_\rho=\dotsc -2, 0, 2 \dotsc}\;\frac{\pi}{4}\,v_\rho K_\rho\, \left(J_\rho+4\,\frac{\!\Phi}{\Phi_0}\right)^2
\ee 
Here, all multiples of $2$ are allowed values for $J_\rho$ by
selection rule \Eq{sel1}. This leads to the various branches shown
in Fig. \ref{branches}. The alternating labels of $\eta=\pm$
reflect the fact that the spin state has to be adjusted according
to selection rule \Eq{sel2} whenever $J_\rho$ is changed by 2.
In the presence of a spin gap, however, this does {\em not} affect
the energy in the limit of \Eq{eq1}, as discussed above. As a
consequence, \Eq{result} has an exact period equal to half a flux
quantum, shown by the lower envelope in Fig. \ref{branches}a). We
note that these findings are in complete agreement with 
those obtained in Ref.
\onlinecite{tJJflux} for the $t$-$J$-$J'$ model.
The amplitude of the ground state energy modulations is apparently given by
\be\label{DeltaE}
  \Delta E = \frac{\pi}{4L} K_\rho v_\rho=\frac{\pi}{4L} v_F^* .
\ee 
where $v_F^*$ is the renormalized Fermi velocity introduced above. 
The corresponding modulations of
the charge current for a given quantum number $J_\rho$ are given by
\be\label{current2}
  \bar{I}_\rho &=& -e\frac{1}{L}\int_0^L dx \,j_\rho(x)\nn
  &=& \frac{2e}{L\pi} v_F^* \,\int_0^L dx
  \left(\partial_x\theta_{\nu}(x)-\frac{2\pi}{L}\frac{\!\Phi}{\Phi_0}\right)\nn
  &=&-\frac{e v_F^*}{L}\left(J_\rho+4\,\frac{\!\Phi}{\Phi_0}\right)\nn
  &=& -c\,\frac{\partial}{\partial\Phi}\,{\cal
  E}\left(\Phi\right)/L.
\ee The current is thus diamagnetic for $-\Phi_0/2<\Phi<\Phi_0/2$
and is given by a sawtooth curve in general which one obtains by
taking the derivative of the envelope in Fig. \ref{branches}a).
The amplitude of the current is given by $\Delta I= e v_F^*/L$, 
which is the same as that of spinless
particles\cite{LOSS}, although the flux period is halved.
Note that this observation is consistent with the notion that the 
charge of the carriers is effectively doubled.

\subsection{The spin gapless case}

To establish the fact that the $hc/2e$ flux period is due to the
presence of a spin gap, it is prudent to demonstrate the change of
flux period when the spin gap collapses. 
First, let us assume $N=4m+2$ as before. Our discussion from the
preceding sections generalizes most easily to the case of a vanishing
spin gap, if we also assume isotropy in the spin sector: In this
case, SU(2) invariance requires the parameter $K_\sigma$
to be unity at the Luttinger liquid fixed point. We will comment 
on the general non-isotropic case below.

For gapless spins, the
operator $H_{1,\perp}$ in \Eq{g-ol2} is irrelevant, and we may expect
to get qualitatively correct results by omitting it. With this
simplification, the spin sector becomes analogous to the charge
sector, and in particular $J_\sigma$ can be regarded as a good
quantum number. The $\eta=+$ spin ground state $\left|+\right>$
then has $J_\sigma=0$, whereas the state $\left|-\right>$ lives in
a degenerate doublet space with $J_\sigma=\pm 2$. This then raises
the corresponding spin energy of the $\left|-\right>$ state by a
term of order $v_\sigma/L$, as shown explicitly in \Eq{HTLdiag}.
As a consequence, the $\eta=-$ branches are shifted upward with
respect to the $\eta=+$ branches (Fig. \ref{branches}b)) which
destroys the $hc/2e$ periodicity of ${\cal E}(\Phi)$.

The 
spin current carrying $\eta=-$ states may (but need not) be
shifted up in energy so much that $\Phi=hc/2e$ ceases to be a
metastable minimum of ${\cal E}(\Phi)$. This is just the case for
a non-interacting system. 
In the case where the uplifting of the $\eta=-$ state is not as large,
$\Phi=hc/2e$ persists to be a metastable minimum in the energy
versus flux curve. The difference between the ground
state energy at $\Phi=0$ and $\Phi=hc/2e$ is thus given by:
\begin{align}\label{shift}
&  \left|E_0(\Phi_0/2)-E_0(0)\right|= \frac{\pi}{L}\min\left(v_\sigma,K_\rho v_\rho\right)\nonumber\\
 &\qquad\qquad  \text{(isotropic spin)}
\end{align} 
Interestingly, \Eq{shift} provides information about the
Luttinger parameters of the {\em spin sector}.
It has long been known that for a Luttinger liquid, the Luttinger
parameters
can be determined from of the  ground state properties.\cite{HAL1}
This technique is often applied to infer the charge Luttinger
parameters, e. g. by calculating the ground state energy as a
function of particle density and magnetic flux.\cite{NOTE2} \Eq{shift} shows
that the same technique may be used to infer spin Luttinger
parameters, 
provided that $v_\sigma<K_\rho v_\rho$
holds. In $SU(2)$ invariant systems, the spinon-velocity $v_\sigma$ 
may thus be obtained. Note that in this case, the $J_\sigma=\pm 2$,
$N_\sigma=0$ states corresponding to the $\eta=\!-$ branches in 
Fig. \ref{branches}b) are degenerate with states having
$J_\sigma=0$, $N_\sigma=\pm 2$, which carry no spin current
but have a net azimuthal spin projection $S_z=\pm 1$. This degeneracy follows from
\Eq{HTLdiag} with $K_\sigma=1$. The latter states, however,
will generally be lower in energy for spin gapless systems {\em without}
SU(2) invariance. This follows because one has $K_\sigma>1$
in this case, since $K_\sigma<1$ would always lead to a spin gap.  
We thus predict that the branches corresponding to the metastable 
minima in Fig. \ref{branches}b), if present, will have a net spin,
rather than a net spin current, in models without SU(2) invariance.
In this case, $v_\sigma$ in \Eq{shift} is to be replaced by
$v_\sigma/K_\sigma$.\cite{NOTE3}

When $N=4m$, 
all the patterns in Fig. \ref{branches} are shifted horizontally
by $hc/2e$. In this case the global minima in Fig.
\ref{branches}b) are located at odd multiples of $hc/2e$. Hence
the function ${\cal E}(\Phi)$ can distinguish the cases $N=4m$ and
$N=4m+2$ in the case of gapless spins, but not in the case of
gapped spins. The same result had also been observed for the
$t$-$J$-$J'$ model.\cite{tJJflux}

We now turn to the case of odd particle number $N=2m+1$. In this
case the selection rules \Eq{sel1} requires  both $N_\sigma$
and $J_\sigma$ to be odd, reflecting the fact that there must be a
dangling spin. (Of course, with a dangling spin the system cannot
have a spin gap.) Now the quantum number
$\eta=\exp\left(-i{\pi}J_\sigma/2\right)$ may take the
values $\pm i$. The two corresponding subsets of the spin state
space are related by the transformation
$J_\sigma\rightarrow -J_\sigma$, which leaves the Hamiltonian
invariant. As a consequence, the spin ground states
$\left|\eta=\pm i\right>$ are exactly degenerate, and an exact
$hc/2e$ periodicity is obtained for the Hamiltonian \eqref{g-ol}
at any system size, regardless of whether $H_1$ is relevant or not.
Also, since $J_\rho$ is now odd as well, the pattern shown in
figure \ref{branches}a) will be shifted horizontally by $hc/4e$.
Hence the global minima of $\cE(\Phi)$ will be located at odd multiples
of $hc/4e$ in this case.

\subsection{Discussion of the results at $T=0$}

The results presented in the preceding section 
are exact for the Hamiltonian \Eq{g-ol}, which is believed
to be the low energy effective theory for all one-dimensional
systems with gapless, linearly dispersing charge degrees of freedom.\cite{NOTE4}
These results may thus be 
expected to be representative for this entire universality class.
To rigorously justify this point, the effects of higher order, 
less relevant operators should be included into the model studied
above. We will not carry out such a detailed analysis here. Rather,
we will point out some expected modifications due to less relevant
operators, and argue for the robustness of the basic results
 derived above by comparing them to special examples of 
microscopic models, where the features of $\cE(\Phi)$ are known 
analytically or numerically.

The flux period of the repulsive Hubbard model was studied in 
Ref. \onlinecite{YUFOW}. These results agree well with our findings
for the spin gapless case. In particular, for odd particle number $N$
the global minima of $\cE(\Phi)$ are at odd multiples of $hc/4e$,
and the flux period is $hc/2e$. While it may seem surprising
that the flux period does not distinguish the spin gapless, odd
particle number case from the spin gapped case (except for the
position of the minima), the microscopic origin of the $hc/2e$ period
is of a rather different nature in the two cases. 
A more subtle effect may demonstrate this:
If one calculates the $\cE(\Phi)$ of {\em free} electrons for odd $N$, 
one indeed finds that $hc/2e$
is the flux period. 
However one also finds that 
there exist corrections to
the $hc/2e$ period at order $1/L$ in ${\cal E}(\Phi)$. 
These corrections are due to the band curvature neglected in the
Hamiltonian \Eq{g-ol}. Similar corrections to ${\cal
E}(\Phi)$ also exist at odd $N$ for the $t$-$J$-$J'$ model. They
can be calculated using the method discussed
in Ref. \onlinecite{tJJflux}.\cite{NOTE5}
Such corrections in powers of $1/L$, however, were not found
in the $t$-$J$-$J'$ model for the spin gapped case, where only
exponentially small corrections were observed. We thus argue
that corrections to the $hc/2e$ flux period generally scale
as $1/L$ in the odd $N$ case, while they are exponentially
small in the spin gapped case. The behavior in the latter case
can  be attributed to the fact that the spin gap generally causes an 
exponentially small sensitivity to boundary conditions in the spin sector. 
This point will be further clarified in Appendix \ref{appA}.

The analysis in the preceding section predicts the ground state to
be unique on the branches of $\cE(\Phi)$ which contain the global 
minima. We expect this to be obeyed by general Hamiltonians.  
However, the four-fold degeneracy which we found between the 
$N_\sigma=0$, $J_\sigma= \pm 2$ and $N_\sigma=\pm 2$, $J_\sigma= 0$
states on the metastable branches
in the isotropic, even $N$, gapless spin case 
is an artifact of our restriction to the Luttinger
Hamiltonian \Eq{HTL}. Rather, the true eigenstates are given
by a triplet and a singlet to be formed from these four states,
giving rise to a small splitting. 
However, except for this effect, the implied degeneracies at the crossings 
between the branches remain valid: Although the conservation 
of $J_\rho$ is approximate once higher order operators are allowed, a
change of $J_\rho$ by 2 implies a change of momentum by
$2k_f$. Hence the states at a branch crossing will not be 
mixed, and the cusps in $\cE(\Phi)$ will remain sharp for general
models.\cite{NOTE6} Finite size studies of the Hubbard
model\cite{FEKULA} show that the patterns displayed in Fig. \ref{branches}
indeed emerge very clearly in numerical simulations carried out at
moderate system size, both for the spin gapped (attractive) and gapless 
(repulsive) case.

We thus conclude that all systems which can be characterized as
Luther-Emery liquids have the $hc/2e$ flux period.
In particular, deviations from the patterns in Fig. \ref{branches} 
such as the appearance of additional minima at higher fractions of 
a flux quantum must be attributed to finite size effects.
Such additional minima at $\Phi_0/n$ are known to occur 
in the large $U$-limit of the Hubbard model \cite{KUS2},
or the small $J$-limit of $t$-$J$-type models \cite{FERRARI}, for
{\em fixed } system size. The criterion for
such finite size effects to disappear is that the amplitude of the
oscillations $\Delta E\sim 1/L$ from \Eq{DeltaE} is small compared to any
other energy scale of the system. In the above cases, the relevant
competing scale is $J\sim t^2/U$. The associated crossover is
clearly observed in Ref. \onlinecite{FERRARI}, where the 
$t$-$J_z$ model is studied: For $t/L\lesssim J_z$, the model
displays the spin gapped behavior shown in Fig. \ref{branches}a).
{This is a consequence of the Ising spin gap of this model.
The similar crossover for the repulsive Hubbard model is shown in Ref.
\onlinecite{YUFOW}, where the pattern of Fig. \ref{branches}b) emerges
(with the necessary shift for $N=4m$).
We note that 
the appearance of local minima separated by half a flux quantum
from the global minima in the repulsive  Hubbard model is 
sometimes interpreted as a sign of pairing.\cite{FEKULA}
We stress, however, that this case {\em does not} meet the criterion 
of a $\Phi_0/2$ flux period as we define it, since a small splitting
of order $t^2/U$ remains between the two types of minima of ${\cal E}(\Phi)$,
which does not vanish as the system size is taken to infinity.}

\subsection{Non-zero temperatures}

Finally, we briefly comment on the expected generalization of our
findings to finite temperature. The behavior stated below can be
verified straightforwardly at the special solvable Luther-Emery
point of the model \Eq{g-ol} (see Appendix \ref{appA}). For $T>0$, we
consider the modulations of the free energy $F(T,\Phi)$ as a
function of flux. The particle number is held fixed, i.e., the
averages are taken in the canonical ensemble. (If the particle
number were allowed to fluctuate, the even/odd effects discussed
above would considerably weaken the sensitivity to flux.)

Under these conditions, the observations made above for the ground
state energy will carry over to the free energy as long as
$T<\Delta E=O(1/L)$. However, the limit of \Eq{eq1} is not to be
taken here, because the amplitude of the free energy modulations
is proportional to $\exp\left(-\text{const } T/\Delta E\right)$
rather than $\Delta E$ when  $T>\Delta E$. In the spin gapped,
even particle number case it remains true that terms violating the
$hc/2e$ periodicity are suppressed by a factor $\sim
\exp\left(-\text{const } \Delta_\sigma L/v_\sigma \right)$, where
$\Delta_\sigma$ is the spin gap. Comparing the two exponential
factors, these $hc/2e$ violating terms will be negligible until
$T$ is of the order of the spin gap. At this temperature,
the amplitude of $\cE(\Phi)$ is already exponentially suppressed,
provided that the system is large enough such that 
$\Delta E \ll \Delta_\sigma$ is satisfied. This again shows
that the $hc/2e$ period will be obeyed as long as $\Delta E$ is the
smallest energy scale (other than temperature) of the system.

We note that the cusps between the branches in Fig. \ref{branches}
will  be smoothened by thermal fluctuations, giving rise
to a finite negative curvature and paramagnetic\cite{LOSS}
effects.

\section{\label{conclusion} Conclusions}

In this paper, we demonstrated that the ground state energy of
Luther-Emery liquids
will generally exhibit an $hc/2e$ flux period.
While this statement holds in a strict sense in the limit of large
system size, finite size deviations are expected to be
exponentially suppressed in the system size. This result had been
anticipated in an earlier work on a particular microscopic
realization of the Luther-Emery liquid, the
$t$-$J$-$J'$-model.\cite{tJJflux} Here, we generalized the result
of Ref. \onlinecite{tJJflux} by showing that the $hc/2e$ flux period is
implied by the 
widely accepted low energy effective theory describing such a
phase. 
As a result, we clarify why the state of the spin sector
impacts upon the flux period when it is commonly believed that in
one dimension spin and charge decouple at low energies. 
An important aspect of our findings is that in systems with even 
particle number $N$, the $hc/2e$ period is
triggered by the spin gap (i.e. pairing) alone and is independent
of whether the superconducting pair-pair correlations are the
dominant long-distance/time correlation function. This may be of
particular value for the correct interpretation of numerical work.
In addition, we have also discussed the expected finite
temperature generalization of our findings.

In Ref. \onlinecite{tJJflux} we stressed the SU(2) invariance of
the model discussed there. This requirement has been relaxed in
the present discussion, where we did not enforce SU(2) invariance.
Instead, only the weaker requirement of a conserved $z$-component
of the spin $S_z=N_\sigma/2$ was found necessary. In the anisotropic 
case, we must also require that the spin gapped ground state has 
$S_z=0$ (see footnote \onlinecite{NOTE4}), which should be automatic in the isotropic
case.

Our findings underline the intuitive notion that every spin gapped
system with linearly dispersing charge modes should share some
features of a superconductor.

\begin{acknowledgments}
\vspace{-.05cm}
This work has been supported by DOE grant DE-AC03-76SF00098.
\end{acknowledgments}

\appendix
\section{\label{appA}Refermionization of the spin Hamiltonian for finite system}

It is well known that the sine-Gordon model  \be\label{Hsigma}
  H_\sigma &&= \frac{v_\sigma}{\pi}\,\int dx\,\bnormal{\left\{K_\sigma\bigl(\partial_x\theta_\sigma(x)\bigr)^2
  +\frac{1}{K_\sigma}\bigl(\partial_x\phi_\sigma(x)\bigr)^2\right\}}\nn&&-
\frac{2g_{1\perp}}{L^2}\,\int
dx\,\bnormal{\cos\bigl(4\phi_\sigma(x)\bigr)}\;, \ee can be mapped
onto the massive Thirring model.\cite{COLEMAN, LE} This mapping
permits a rather direct demonstration of the exponentially small
energy difference between the $\eta=+$ and $\eta=-$ states
discussed in section \ref{gapped}. In addition, when $K_\s=1/2$, i.e. at the
Luther-Emery point, the massive Thirring model reduces to a
massive free fermion Hamiltonian which allows the exact calculation of
various physical quantities.
In particular finite temperature results can be obtained at the
Luther-Emery point easily.

In the notation established in section \ref{bosonize}, the mapping
onto the massive Thirring model can be performed by the
introduction of the following spinless fermion operators:
\begin{equation}\label{referm}
\begin{split}
   \tilde \psi_r^\dagger (x)\; &= \;\frac{e^{\,i\frac{\pi}{L}rx}}{\sqrt{L}} \;\tilde A_r\;\bnormal{e^{\,i\bigl(2r\phi_\sigma(x)+\theta_\sigma(x)\bigr)}}\\
   \text{where }\qquad \tilde A_r &= e^{\,i \frac{\pi}{2}\left(\frac{1}{2}rN_\sigma-\frac{1}{4}J_\sigma\right)}
\end{split}
\end{equation}
Here, the symbol $\bnormal{\,}$ denotes a normal ordering
convention analogous to that defined in section \ref{bosonize},
but where $\varphi_{r,s}$ is replaced by the field
\begin{equation}\label{phitilde}
\begin{split}
   \tilde \varphi_r(x)&=\frac{1}{4}\,\sum_s\,s\,\left({3}\varphi_{r,s}(x)-\varphi_{-r,s}^\dagger(x)\right)\;\\
{\overline{\tilde\varphi}}_r&=\frac{1}{4}\,\sum_s\,s\,\left({3}\overline{\varphi}_{r,s}-\overline{\varphi}_{-r,s}\right)=2r\overline\phi_\sigma+\overline\theta_\sigma\quad.
\end{split}
\end{equation}

{ 
It is interesting to note that the operators $ \tilde \psi_r (x)$,
in terms of which the spin Hamiltonian \Eq{Hsigma} is best analyzed,
lead out of the physical Hilbert space $\cH_{phys}$: Apart from affecting the
spin current quantum number $J_\sigma$, they also change the total number of
net excited spins, $N_\sigma$, by $1$. In the physical Hilbert space, such a change must
always go along with a change of charge quantum numbers, which are not 
affected by $ \tilde \psi_r (x)$. It is quite natural that the action
of a single ``fractionalized'' operator such as  $ \tilde \psi_r (x)$ will
lead out of the space of physical states. At this point the larger
space $\cH_{frac}$ discussed in Section \ref{bosonize} becomes indispensable,
as it allows us to define operators such as $ \tilde \psi_r (x)$
in the first place. It is clear, however, that these operators enter
the Hamiltonian only in appropriate pairs, which  leave the physical 
subspace invariant.
}

Using standard methods reviewed in Ref. \onlinecite{DELFT}, it is
straightforward to show that the field defined in \Eq{referm}
satisfies the required anticommutation relations. The additional
factor $\exp(i\pi rx/L)$ in \Eq{referm} will be commented on %
below. For now we note that it gives rise to
the following boundary conditions for the spinless fermion fields:
\be\label{boundary}
  \tilde\psi_r(x+L)\,=\,e^{\,i\pi\left(N_\sigma+J_\sigma/2-1\right)}\,\tilde \psi_r(x)
\ee

As is relevant to section \ref{proof}, in the following we will concentrate
on the case $N_\sigma=0$ and $J_\sigma$ even. \Eq{boundary} then
tells us that the sector $\eta=-$ is represented by fermions
obeying periodic boundary conditions, and the $\eta=+$ sector by
fermions obeying antiperiodic boundary conditions. Note that
without the additional twist in \Eq{referm}, it would have been
vice versa.

Using the methods discussed in Ref. \onlinecite{DELFT}, one may
now show the equivalence of \Eq{Hsigma} and \be H_{mtm}&=&
\sum_r\int_0^L dx~\left(-irv\right):
\tilde\psi_r^\dagger(x)\partial_x\tilde\psi_r(x):\nn&+&\sum_r\int_0^L
dx~m(i)^r~:\tilde\psi_r^\dagger(x)\tilde\psi_{-r}(x):\nn
&+&g\sum_r\int_0^L
dx~\bnormal{\tilde\psi_r^\dagger(x)\tilde\psi_r(x)}
\bnormal{\tilde\psi_{-r}^\dagger(x)\tilde\psi_{-r}(x)}\nn&\equiv&
H_\sigma+\text{const,}\label{HMTM}\ee where\cite{NOTE8} 
\be\label{params}
v&&=\frac{v_\sigma}{4}\left(\frac{1}{K_\sigma}+4K_\sigma\right)\nonumber\\
g&&=\frac{\pi v_\sigma}{4}\left(\frac{1}{K_\sigma}-4K_\sigma\right)\\
m&&=\frac{|g_{1\perp}|}{2\pi\alpha}\nonumber \ee Here, the symbol
$\bnormal{\,}$ is as defined in \Eq{fnormal}, but the vacuum
state is now the $N_\sigma=J_\sigma=0$ state that is annihilated
by the field \Eq{phitilde}. 

\begin{figure}[t]
\samepage
\begin{center}
\vspace{.0cm}
\includegraphics[height=2in,width=2in]{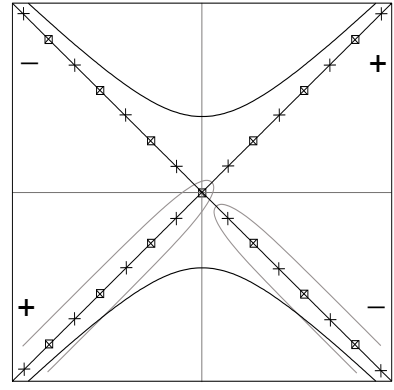}
\caption{\label{ref} Ground state of the fermion Hamiltonian \Eq{HMTM} for $g=0$.
The linear branches indicate the massless case where both $N_\sigma$ and $J_\sigma$
are good quantum numbers. Crosses indicate allowed momenta in the $\eta=+$ sector,
boxes those in the $\eta=-$ sector. The encircled areas indicate the occupied
states on each branch for the ground states in the $N_\sigma=0$, $J_\sigma=0$
sector (crosses), as well as  the $N_\sigma=0$, $J_\sigma=+2$ sector (boxes). The
 latter state is degenerate with the $N_\sigma=0$, $J_\sigma=-2$ ground state,
where the $k=0$ mode is occupied on the $r=-$ branch instead. For non-zero mass and
 $N_\sigma=0$, a gap opens up and the upper band remains empty while the lower band
is completely filled, where the allowed momenta depend on $\eta$ as shown.}
\end{center}
\vspace{0cm}
\end{figure}

When 
$K_\sigma<1$ the sine-Gordon model is massive, and the physics of
\Eq{HMTM} is given by massive spinless fermions. 
In particular, the fermionic interaction $g$ vanishes for 
$K_\sigma=1/2$, and the Hamiltonian \Eq{HMTM} becomes that of a
massive free fermion model. This is the special point identified
by Luther and Emery.\cite{LE} At the Luther-Emery point the
fermion dispersion relation is given by $\epsilon(k)=\sqrt{(vk)^2+m^2}$.
In a range of $K_\sigma$ values around $1/2$, the $g$-term only gives rise to
quantitative corrections. 

We now return to the factor $\exp(i\pi rx/L)$ in \Eq{referm}
(or the boundary conditions \Eq{boundary}) and show that with
this factor (or the boundary condition specified in
\Eq{boundary}) the boson and fermion theories are consistent. To
illustrate that we compare the ground state degeneracies in the
non-interacting massless case for the bosonic and fermionic
theories, i. e., we let $K_\sigma=1/2$ and $ g_{1\perp}=0$, which
results in $m=g=0$.

In the absence of $g_{1\perp}$ both $N_\s$ and $J_\s$ are
good quantum numbers. Let us denote the ground state in the
$N_\s,J_\s$ sector by $\left|N_\sigma,J_\sigma\right>$. From
\Eq{HTLdiag} we recall that  $\left|0,0\right>$ is the
non-degenerate global ground state of the spin sector, whereas the
states $\left|0,2\right>$, $\left|0,-2\right>$ form a degenerate
doublet. That this also holds in the fermionic representations of
the model is just achieved by the boundary conditions
\Eq{boundary} (see Fig. \ref{ref} and the caption). 
Note that since the
fermions \Eq{referm} are {\em derived} in terms of bosons, the
fermion occupancies claimed below Fig. \ref{ref} follow by
derivation, not by definition. In fact one may easily evaluate \be
\left<N_\sigma,J_\sigma|\,\tilde c_r^\dagger(k)\tilde
c_r(k)\,|N_\sigma,J_\sigma\right>,\ee where \be &&\tilde
c_r^\dagger(k)\equiv\frac{1}{\sqrt{L}}\int_0^Ldx\,
e^{\,ikx}\,\tilde\psi_r^\dagger(x)\nn&&
k=\frac{2\pi}{L}n+\frac{\pi}{L}\left(N_\sigma+J_\sigma/2-1\right)
\ee by plugging in \Eq{referm}, and verify that the occupancies
identified for the various states in Fig. \ref{ref} are
correct.\cite{NOTE7}

The spin sector of the model is now represented in terms
of fermions satisfying a conventional boundary condition given by 
\Eq{boundary}.
The difference between the ground state
energy for $\eta=+$ and $\eta=-$  thus becomes 
the change in the 
fermion ground state energy induced by a change of the boundary condition,
or equivalently the modulation of the fermion ground state energy
caused by an AB flux. When the sine-Gordon model is massive, the
fermions form an insulating state. Then, the sensitivity of their
ground state energy to the boundary condition will vanish
exponentially with the system size, as is well known from the 
general arguments given by Kohn\cite{KOHN} and Thouless\cite{THOULESS2}.
The greatest advantage of the refermionization occurs at the
Luther-Emery point. For in that case, a non-trivial interacting
bosonic theory is mapped onto a free fermion theory. 
In particular, at the Luther-Emery point
$g=0$ one obtains from a direct calculation that the energy
difference between the ground states for $\eta=+$ and $\eta=-$
vanishes as $m\exp(-mL/v)$, as we claimed earlier.\\

\section{\label{appB}Number and phase variables for continuum and lattice Hilbert spaces}

The Hilbert space of the Luttinger Hamiltonian \Eq{HTL},
denoted as the ``physical'' Hilbert space $\cH_{phys}$ in the
bulk of the paper, can be decomposed as
\begin{equation}\label{Hphys}
  \cH_{phys}=\bigotimes_{r,s}\, \cH^{N_{r,s}}_{l}\otimes\cH^{b_{r,s}}
\end{equation}
Here, the spaces $\cH^{b_{r,s}}$ contain all the degrees of freedom associated with the bosonic excitation spectrum, whereas $\cH^{N_{r,s}}_{l}$ contains the degrees of freedom of the operator $N_{r,s}$. The subscript ``$l$'' stands for ``lattice'' and reminds us of the discrete nature of $N_{r,s}$ in the physical Hilbert space: The Hilbert space basis of $\cH^{N_{r,s}}_{l}$ is given by a set of non-degenerate eigenstates of $N_{r,s}$, whose spectrum consists of all integer numbers. 

In the process of bosonization, however, we introduce new linear combinations $N_\nu$, $J_\nu$  of the $N_{r,s}$ (\Eq{qnumbers}). The spectrum of these new operators is likewise integer, yet not all possible combinations of integer eigenvalues are physically allowed. This ``residual coupling'' is not evident from commutation relations, since the operators  $N_\nu$, $J_\nu$ all commute as the $N_{r,s}$ do. Hence, once we bosonize an enlarged Hilbert space becomes more natural, where the eigenvalues of the operators  $N_\nu$, $J_\nu$ are independent. This is the Hilbert space $\cH_{frac}$ of ``fractional'' excitations. The physical subspace $\cH_{phys}$ is then characterized by the fact that the selection rules \Eq{sel} are satisfied. It is clear that in $\cH_{frac}$, the spectrum of the $N_{r,s}$ must also contain certain fractional values. Formally, we find it convenient to introduce an even larger Hilbert space $\cH$, where the spectrum of the $N_{r,s}$ is {\em continuous}. The benefit of this is that the conjugate phase $\overline \varphi_{r,s}$ of these operators then becomes meaningful. This, in turn, allows us to construct unitary ``ladder operators'' which change the eigenvalue of the $N_{r,s}$ by {\em arbitrary} amounts. This formalism is of particular advantage in Appendix \ref{appA}, where fractionalized spin fermion operators are constructed. Below we present some fine details of this embedding of $\cH_{phys}$ into the larger space $\cH$.

For this purpose let us consider a single operator $\hat N$ and its conjugate variable $\hat \varphi$ such that
\begin{equation}
  \label{commB}
  \left[\hat\varphi,\hat N\right]=i
\end{equation}
holds. An analogy to the quantum mechanics of a point particle moving in one dimension is obtained if we identify $\hat N\equiv \hat x$ and $\hat \varphi\equiv -\hat p$, where $\hat x$ and $\hat p$ are the coordinate and the momentum of the particle. In this context, it is familiar how to construct a Hilbert space $\cH^N$ such that $\hat N$ and $\hat \varphi$ are well defined on a dense set, \Eq{commB} is satisfied and the spectrum of $\hat N$ is unbounded: It is the Hilbert space of square integrable functions of the variable $N$. From the commutation relation \Eq{commB}, it is clear that the spectrum of both $\hat N$ and $\hat \varphi$ has to be continuous and unbounded. In particular, we can construct shift operators $\exp (i\hat\varphi a)$ satisfying
\begin{equation}
  \label{commB2}
 \left[\hat N,e^{i\hat\varphi a} \right]=a \, e^{i\hat\varphi a}, 
\end{equation}
which shift the value of $\hat N$ by an arbitrary amount $a$.

We note that, as is familiar from the point particle analogy, the ``position'' and ``momentum'' eigenkets $\ket{N}$ and $\ket{\varphi}$ are not strictly contained in the Hilbert space $\cH^N$ of ``proper'' vectors, but are ``generalized'' states in the usual sense. Here, we will not attempt to introduce a different notation for proper and generalized states, nor for the proper Hilbert space and its extension containing generalized states. We refer the reader to Ref. \onlinecite{BOHM} and references therein for details, and simply note that the kets $\ket{N}$ and $\ket{\phi}$ satisfy
\begin{equation}
  \begin{split}
     \braket{N'}{N}=\delta(N'-N)\;,\;\braket{\varphi'}{\varphi}=2\pi\delta(\varphi'-\varphi)\\
\ket{\varphi}=\int dN\,e^{-i\varphi N}\ket{N}\,,\,\ket{N}=\frac{1}{2\pi}\,\int d\varphi\,e^{i\varphi N}\ket{\varphi}
  \end{split}
\end{equation}

Suppose now that a physical problem is defined on a subspace
$\cH^N_{l}$, which is given by the discrete ``lattice'' 
represented by the eigenkets $\ket{N}$ for integer $N$.
Since these kets from a countable Hilbert space basis in 
$\cH^N_{l}$, it is natural and convenient to introduce
a new scalar product on $\cH^N_{l}$ via:
\begin{equation}\label{scalphy}
  \braketphy{N'}{N}=\delta_{N',N}\;.
\end{equation}
This differs from the scalar product in $\cH^N$ only by an infinite 
multiplicative factor. In the above, $\ketphy{N}$ denotes the same vector
as $\ket{N}$, but endowed with a different scalar product.
\Eq{scalphy} means that the $\ket{N}$ for integer
$N$ become a complete orthogonal set of {\em proper} vectors within 
$\cH^N_{l}$. Within $\cH^N_{l}$, one may now define a ``crystal momentum''
operator $\hat\varphi_{l}$, whose eigenkets are defined to be
\begin{equation}\label{xtal}
  \ketphy{\varphi}=\sum_{N\in \mathbb{Z}} e^{-i\varphi N}\ketphy{N}\quad.
\end{equation}
One observes that these eigenkets are periodic in $\varphi$ with period $2\pi$, hence for definiteness the eigenvalues must be restricted to lie within the ``Brillouin zone'' $(-\pi,\pi]$, where
\begin{equation}\label{scalphi}
 \braketphy{\varphi'}{\varphi}=2\pi\,\delta(\varphi'-\varphi)\,,\quad\varphi',\varphi\in (-\pi,\pi]
\end{equation}
holds. If we now denote the restriction of $\hat N$ to $\cH^N_{l}$ by $\hat N_{l}$, we find that the commutator $[\hat\varphi_{l},\hat N_{l}]$ is {\em not quite} analogous to \Eq{commB}. This has been examined in detail by Sch\"onhammer.\cite{SCHONHAMMER1,SCHONHAMMER2} However, for the applications we have in mind here, this difference never matters. This is so since all physical observables, including the Hamiltonian, depend on $\hat\varphi_{l}$ only via integer powers of $\exp(i\hat\varphi_{l})$, and since the equations

\begin{equation}
  e^{i\hat\varphi}\ket{N}=\ket{N+1},\quad e^{i\hat\varphi_{l}}\ketphy{N}=\ketphy{N+1}
\end{equation}

\noindent hold. Hence $\exp(i\hat\varphi_{l})$ and $\exp(i\hat\varphi)$ {\em act identically} on  $\cH^N_{l}$, and we may express all observables in terms of {\em either} of these operators. By means of \Eq{xtal}, the kets $\ketphy{\varphi}\in\cH^N_{l}$ are identified with the following kets in $\cH^N$:
\begin{equation}\label{phiphy}
  \ketphy{\varphi}\,\equiv\, \sum_{N\in \mathbb{Z}} e^{-i\varphi N}\ket{N}
=\sum_{n\in \mathbb{Z}} \ket{\varphi+2\pi n}
\end{equation}
Note that the norm of the right hand side with the scalar product in $\cH^N$ formally computes to $\infty\times\delta(0)$, which by comparison with \Eq{scalphi} is larger by an infinite multiplicative constant than $\braketphy{\varphi}{\varphi}$. Recall that the same relation also holds for $\braket{N}{N}$ and $\braketphy{N}{N}$ by definition of the scalar product in  $\cH^N_{l}$. \Eq{phiphy} makes it clear that the kets  $\ketphy{\varphi}\in\cH^N_{l}$ are {\em not} to be identified with the kets $\ket{\varphi}\in\cH^N$. The latter are not periodic in $\varphi$, and cannot be constructed solely from kets $\ket{N}$ with integer $N$. However,
\begin{equation}
  \braket{\varphi+2\pi n}{N}=\braketphy{\varphi}{N} \qquad \forall n\in\mathbb{Z}
\end{equation}
holds. Thus whenever it is clear from the context that we are working in $\cH^N_{l}$, we may drop all labels ``$l$'', keeping the periodicity of the states $\ket{\varphi}\in\cH^N_{l}$ in mind.

We can now define the Hilbert space
$\cH$ introduced in the main part of the paper as
\begin{equation}
      \cH=\bigotimes_{r,s}\, \cH^{N_{r,s}}\otimes\cH^{b_{r,s}}\quad.
\end{equation}
The advantage of embedding $\cH_{phys}$ into the space $\cH$ is that 
now the space $H_{frac}$, satisfying $\cH_{phys}\subset\cH_{frac}\subset\cH$,
can be generated easily through the action of the operators $\exp(i\overline\theta_\nu)$, $\exp(i\overline\phi_\nu)$  on  $\cH_{phys}$. The bookkeeping 
is greatly simplified by the simple commutation relation of $\overline \varphi_{r,s}$ and $N_{r,s}$, \Eq{comm0}, valid on $\cH$. We finally note that $\cH_{frac}$ can be written as a product analogous to \Eq{Hphys}, involving a space containing bosonic degrees of freedom and four discrete ``lattice'' spaces containing the degrees of freedom of the quantum numbers $N_\nu$, $J_\nu$. All the above therefore holds in an analogous way for $\cH_{frac}$, and in particular lattice versions of the phase operators $\overline \phi_\nu$, $\overline \theta_\nu$ can be constructed if desired.

In Ref. \onlinecite{SCHONHAMMER2}, it has been noted that the use of the canonical commutation relations Eqs. \eqref{comm0}, \eqref{commB} in constructing the Klein factors does indeed yield correct results in bosonization, even though these relations cannot be rigorously justified for operators that are restricted to a discrete space. We believe that the embedding procedure discussed here provides a proper explanation for this observation.


\begin{thebibliography}{38}
\bibitem[{NOT({\natexlab{a}})}]{NOTE1}
\bibinfo{note}{In general, any linear combination of $\overline\phi_\nu$ and
  $\overline\theta_\nu$ may be exponentiated which can be expressed in terms of
  the $N_{r,s}$ with integer coefficients.}
\bibitem[{NOT({\natexlab{b}})}]{NOTE9}
\bibinfo{note}{Note that this reinterpretation of $\overline\phi_\sigma$ leaves
  the cosine-interaction invariant, as discussed in Appendix \ref{appB}.}
\bibitem[{NOT({\natexlab{c}})}]{NOTE2}
\bibinfo{note}{In principle, this method is not limited to the charge sector
  though, if one formally assigns opposite electric ``charges'' to upspin and
  downspin particles when coupling to flux.}
\bibitem[{NOT({\natexlab{d}})}]{NOTE3}
\bibinfo{note}{If one restricts the Hamiltonian to the $S_z=0$ sector, one must
  replace $v_\sigma$ by $v_\sigma K_\sigma$ instead.}
\bibitem[{NOT({\natexlab{e}})}]{NOTE4}
\bibinfo{note}{Systems with a macroscopic magnetization in their ground state,
  which may also have a spin gap in the case of anisotropy, should be excluded
  from this statement.}
\bibitem[{NOT({\natexlab{f}})}]{NOTE5}
\bibinfo{note}{In the language of Ref. \onlinecite{tJJflux}, the spin momentum
  $K$ that minimizes the energy of the spin sector approaches $K=\pm \pi/2$ for
  large odd $N$. However, since a finite odd length spin chain cannot precisely
  have this momentum, the minima of ${\cal E}(\Phi)$ are slightly displaced
  from odd multiples of $\Phi_0/4$ by alternating amounts. This gives rise to
  the same $1/L$ corrections to the $\Phi_0/2$ flux period that are also
  observed for non-interacting, odd $N$ systems.}
\bibitem[{NOT({\natexlab{g}})}]{NOTE6}
\bibinfo{note}{Except at half filling, if $\cE(\Phi)$ only consist of branches
  whose $J_\rho$ values differ by multiples of 4. In this case, one may not
  expect sharp cusps even in the absence of a charge gap.}
\bibitem[{NOT({\natexlab{h}})}]{NOTE8}
\bibinfo{note}{The appearance of the infinitesimal $\alpha$ in the denominator
  of the mass parameter is a sign that the cosine interaction in \Eq{Hsigma} is
  not properly regularized by simply normal ordering in the {"bare"} boson
  field $\varphi_{r,s}$. The problem can be circumvented by imposing the
  {"new"} normal ordering prescription defined below \Eq{referm} for the spin
  sector from the start. The infinitesimal $\alpha$ would then be absent from
  the definition of the mass parameter. However, the expression \Eq{params} for
  the mass parameter is standard in the literature (c.f. Ref. \onlinecite{LE}),
  and $v_F/\alpha$ can be interpreted as the bandwidth of an underlying
  microscopic system.}
\bibitem[{NOT({\natexlab{i}})}]{NOTE7}
\bibinfo{note}{Note that all the states $\left|N_\sigma,J_\sigma\right>$
  defined in the text are just annihilated by the $q\neq 0$ part of the fields
  $\tilde\varphi_r(x)$.}
\end{thebibliography}

\end{document}